# Optimisation and Performance Computation of a Phase Frequency Detector Module for IoT Devices

**Md. Shahriar Khan Hemel[1], Mamun Bin Ibne Reaz[2], Sawal Hamid Bin Md Ali[1], Mohammad Arif Sobhan Bhuiyan[3,*] and Mahdi H. Miraz[4,5,6,*]**

[1]Department of Electrical, Electronic and Systems Engineering, Universiti Kebangsaan Malaysia, Malaysia
P114543@siswa.ukm.edu.my, sawal@ukm.edu.my
[2]Department of Electrical and Electronic Engineering, Independent University, Bangladesh
mamun.reaz@iub.edu.bd
[3]School of Electrical Engineering and Artificial Intelligence, Xiamen University Malaysia, Malaysia
arifsobhan.bhuiyan@xmu.edu.my
[4]School of Computing and Data Science, Xiamen University Malaysia, Malaysia
m.miraz@ieee.org
[5]Faculty of Arts, Science and Technology, Wrexham University, UK
m.miraz@ieee.org
[6]Faculty of Computing, Engineering and Science, University of South Wales, UK
m.miraz@ieee.org
*Correspondence: m.miraz@ieee.org; arifsobhan.bhuiyan@xmu.edu.my

**Abstract: The Internet of Things (IoT) is pivotal in transforming the way we live and interact with our surroundings. To cope with the advancement in technologies, it is vital to acquire accuracy with the speed. A phase frequency detector (PFD) is a critical device to regulate and provide accurate frequency in IoT devices. Designing a PFD poses challenges in achieving precise phase detection, minimising dead zones, optimising power consumption, and ensuring robust performance across various operational frequencies, necessitating complex engineering and innovative solutions. This study delves into optimising a PFD circuit, designed using 90 nm standard CMOS technology, aiming to achieve superior operational frequencies. An efficient and high-frequency PFD design is crafted and analysed using cadence virtuoso. The study focused on investigating the impact of optimising PFD design. With the optimised PFD, an operational frequency of 5 GHz has been achieved, along with a power consumption of only 29 μW. The dead zone of the PFD was only 25 ps.**

---

## 1. Introduction

The Internet of Things (IoT) is crucial for enabling smart, connected ecosystems that enhance efficiency and convenience in various aspects of our lives. It empowers businesses, healthcare, transportation and homes by facilitating real-time data exchange as well as automation [1-2]. To cope with the technological advancement, devices with better speed, accuracy and low power consumption have become integral constituents of the IoT ecosystems.

IoT transceiver refers to a specialised communication device designed for IoT applications, allowing these devices to exchange data with other devices or networks. IoT transceivers are characterised by their ability to wirelessly transmit and receive data, enabling seamless communication amongst various IoT devices and the central network or cloud platform. They play a pivotal role in enabling smart homes, industrial automation, healthcare systems and other IoT-enabled ecosystems, allowing diverse devices to interact and share information, fostering the growth of the Internet of Things.

For seamless communication amongst IoT devices, researchers employ phase-locked loop (PLL) in IoT transceivers for generating a regulated low noise frequency [3]. This synchronisation is important for maintaining the integrity of transmitted data, minimising signal interference and ensuring reliable reception. PLLs also facilitate frequency synthesis, enabling transceivers to operate on various channels and adapt to different communication standards.

The operation of PLL begins with the PFD, which compares the phases of the reference and feedback signals. Any phase difference results in an error signal, driving the charge pump (CP) [4-5]. The CP converts this error signal into a continuous voltage, proportionally adjusting the VCO's frequency. The LF then filters this voltage, removing high-frequency noise, and delivers a smooth control voltage to the VCO. Consequently, the VCO tunes its frequency to align with the reference signal, minimising the phase error [6]. The FD further divides the VCO output, providing a divided output signal. The feedback loop continuously refines the VCO's frequency until it precisely matches the reference signal [7]. This synchronised output finds applications in numerous fields, including telecommunications, where stable, synchronised signals are vital for reliable data transmission.

A phase frequency detector (PFD) is a fundamental component in digital and analogue systems [8-9], primarily used in phase-locked loops (PLLs) and frequency synthesisers [10]. Its primary function is to compare the phase difference between the two input signals and generate an output voltage [11]. In the context of PLLs, the PFD compares the phase of the reference signal and the feedback signal, enabling the PLL to adjust the output frequency in order to match the desired frequency. The accuracy and speed of this phase comparison process are vital for stable and efficient operation of various electronic systems. Therefore, the PFD is considered as a critical element in applications such as frequency synthesis, clock recovery and communication systems. Figure 1 illustrates the block diagram of a typical PLL architecture [12].

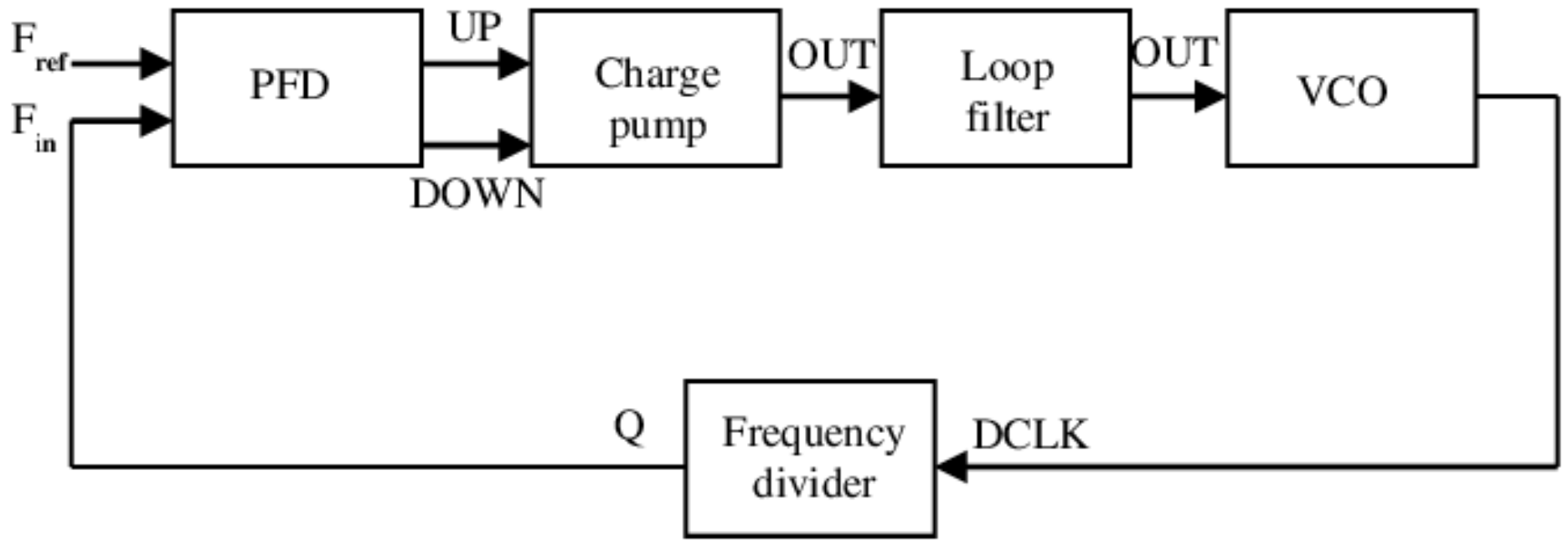

**Figure 1.** Block diagram of a phase locked loop [13]

Complementary metal-oxide-semiconductor (CMOS) technology has long been the cornerstone of modern integrated circuits due to its low power consumption, high noise immunity and cost-effectiveness. Its ability to combine both digital and analogue functions on a single chip makes it highly versatile. Alternatives to CMOS, such as bipolar junction transistor (BJT) and N-type metal-oxide-semiconductor (NMOS) () technology, offer different advantages but often lack the energy efficiency and scalability that CMOS provides. CMOS technology's superiority lies in its minimal power requirements, making it ideal for portable devices, smartphones and battery-operated gadgets. Additionally, CMOS circuits generate less heat, ensuring stable operation even in high-density devices. Its compatibility with digital and analogue components within the same chip simplifies complex designs, reducing manufacturing costs. These combined advantages make CMOS the preferred choice in most electronic applications, especially where energy efficiency, cost-effectiveness and integration complexity are paramount considerations. Therefore, our research has opted for implementation of optimisation technique in CMOS.

Over the decades, researchers have been trying to employ various optimisation techniques to escalate the performance of PFD. In 2015, Anushkannan [14] used pass transistors to optimise power consumption, however, the operational frequency was very low. In 2016, Gholami [15] used transmission gate-based PFD, which achieved relatively higher operational frequency, but the power consumption was excessively high. In the subsequent year, Gholami [16] further modified the conventional PFD, which was able to achieve satisfactory operation frequency, however, the power consumption still remained very high. In 2019, Sofimowloodi [17] designed a symmetric PFD with relatively high operational frequency and low power consumption, but it was not still satisfactory in terms of operational frequency and power consumption. In this study, we have used cadence virtuoso as a computational tool to study the effect of varying MOSFET width, to obtain high operational frequency in a PFD topology, which is a key component in frequency synthesisers. Using this tool, we evaluated the optimal width to obtain the high operational frequency and optimise the power consumption of the PFD.

## 2. Methodology

The inputs to a PFD are typically a reference signal and a feedback signal, often originating from an external source and an internal signal, respectively. These input signals are compared within the PFD and based on their phase relationship, the PFD generates output signals. Common output signals include UP and DOWN signals. When the reference signal leads the feedback signal, the PFD generates UP pulses and when the feedback signal leads the reference signal, it produces DOWN pulses. The UP and DOWN signals are utilised by the charge pump circuit to generate a control voltage, which is then filtered and used to adjust the frequency of a voltage-controlled oscillator (VCO). Through this feedback loop, the PFD ensures that the phase and frequency of the output signal align with the reference signal, allowing for precise synchronisation and stable operation in various electronic applications.

The performance parameters of CMOS PFDs are critical aspects that influence the overall efficiency of integrated circuits. The operational frequency, indicating the speed at which the PFD can function, is a primary parameter. A lower operational frequency can limit the device's suitability and adoption in real-time applications. Dead zone, on the other hand, represents the range within which the PFD cannot distinguish between the reference and the feedback signals. Minimising the dead zone ensures accurate phase detection. Power consumption is crucial, particularly in portable devices, as it directly affects battery life. Additionally, the die area, or the physical space occupied by the PFD on the chip, is a significant consideration, especially in compact devices where space is limited. The choice of CMOS technology, such as 90 nm or 45 nm, impacts the overall performance, power efficiency and cost of the PFD. Optimising these parameters is essential for achieving a balance amongst speed, accuracy, power efficiency and space utilisation in CMOS PFD designs.

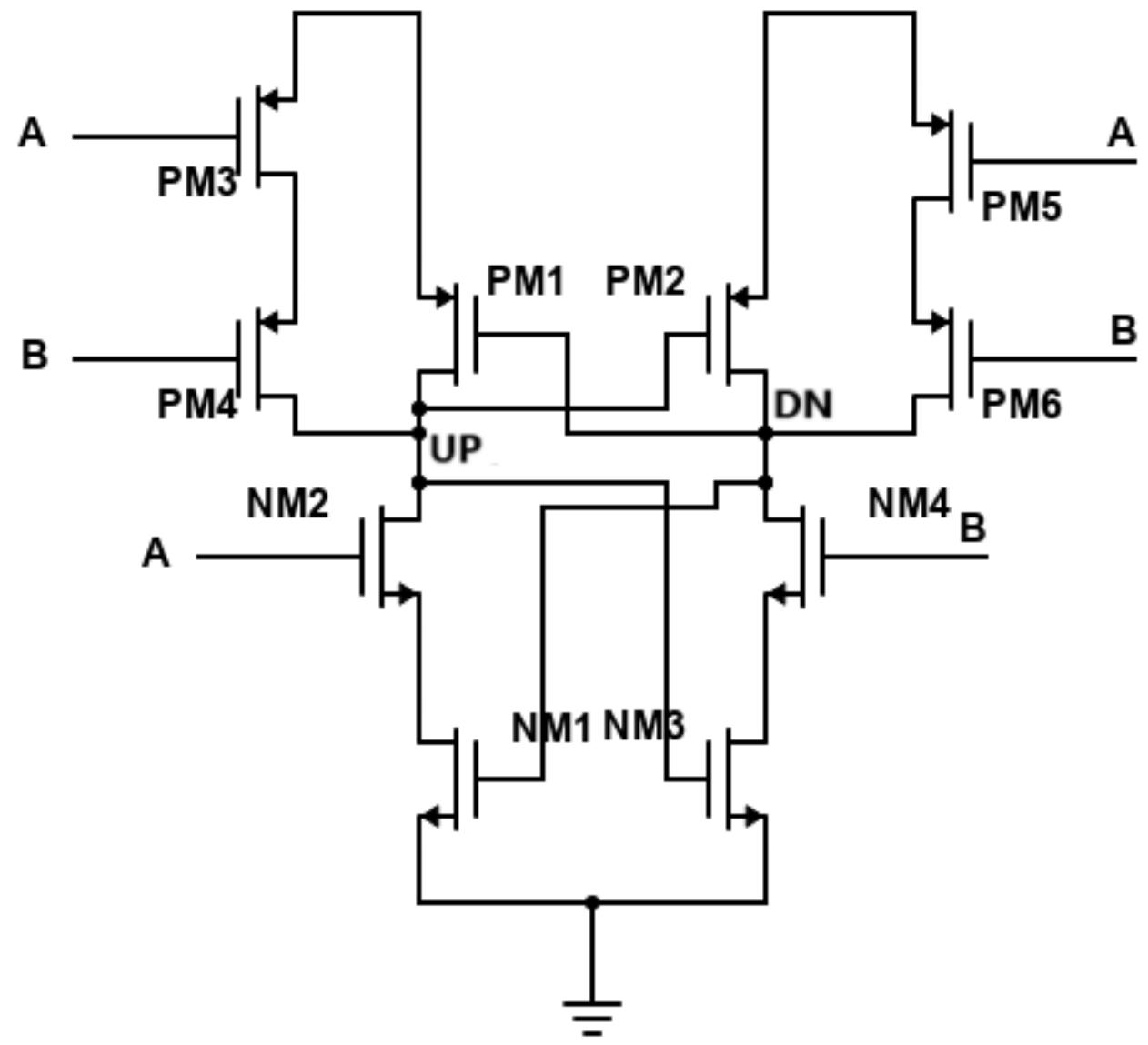

**Figure 2.** Schematic of the core of the PFD

Figure 2 presents the schematic diagram of the core of the PFD which consists of 10 MOSFETS. Amongst them, four are NMOS (NM1 – NM4) and six are PMOS (PM1 – PM6). The NM1 – NM4 MOSFETs are responsible for detecting phase change in the inputs, where PM1 and PM2 have been employed to hold the results. The rest of the MOSFETs (PM3 – PM6) have been introduced to restore the PFD to its default state. Initially, A and B both are 0, as a result UP and DN node charges up to 1. If the input signal A leads B, then UP discharges with the rising edge of the A. As long as A is high, the input signal B does not have any effect on the outputs. As a result, during this period UP remains low whereas DN remains high. On the other hand, if the input signal B leads A the DN node discharges. As long as input B is active, the rising edge of input A doesn't affect the outputs. As a result, during this period, the output UP remains high and DN remains low. One of the major benefits of this design is that both the outputs, i.e., UP and DN, cannot be active at the same time, which extensively reduces the power consumption of the design. In the final design presented in Figure 3, PM4 – PM7 and NM4 - NM7 have been introduced. PM4, PM5, NM4, NM5 and PM6, PM 7, NM6, NM7 are employed as NOR gates to restore the output signals.

Optimising CMOS MOSFET size in VLSI design is pivotal for achieving optimal performance in modern electronic devices. The size of the MOSFETs directly impacts the speed and power efficiency of the integrated circuits. Smaller MOSFETs switch faster, leading to higher operational speeds and improved overall performance. Additionally, reduced transistor size lowers gate capacitance, decreasing power consumption during switching events and thus enhancing energy efficiency which is a major factor for battery-powered devices. Moreover, optimising MOSFET size allows for higher component density on a silicon wafer, enabling the integration of more functions into a smaller space. This miniaturisation not only enhances the device's overall performance but also reduces manufacturing costs and promotes the development of compact, powerful and energy-efficient electronic gadgets that power our modern digital world.

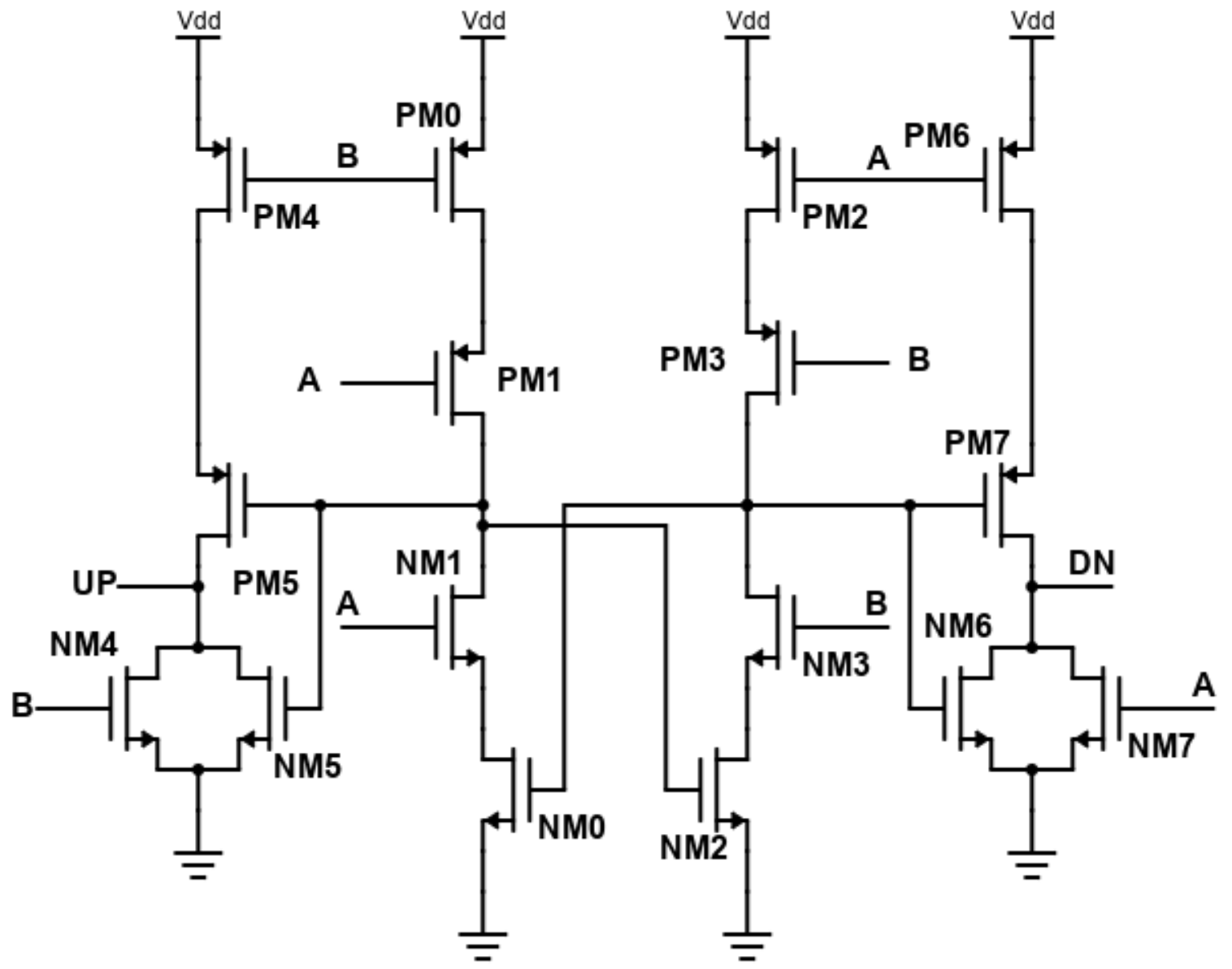

**Figure 3.** Schematic of the proposed PFD

The operational frequency of the PFD circuit is intricately linked to the rise and fall times of the MOSFET. A lower rise and fall time signifies higher switching speeds for the MOSFET and with lower rise and fall time, MOSFETs can operate at higher frequencies. This increased speed, in turn, enables the circuit to operate at a higher frequency. A broader MOSFET width leads to a shorter rise time and consequently, a higher switching speed. However, this enhanced speed comes at a cost, i.e., greater power consumption for the circuit. Hence, a delicate balance must be struck between the switching speed and the power consumption. This study employs Cadence Virtuoso's parametric analysis alongside the 90 nm CMOS technology to meticulously optimise the MOSFET width. The aim is to identify an optimal point where the PFD can operate at a higher frequency without incurring excessive power consumption, navigating the intricate trade-off between the speed and the energy efficiency.

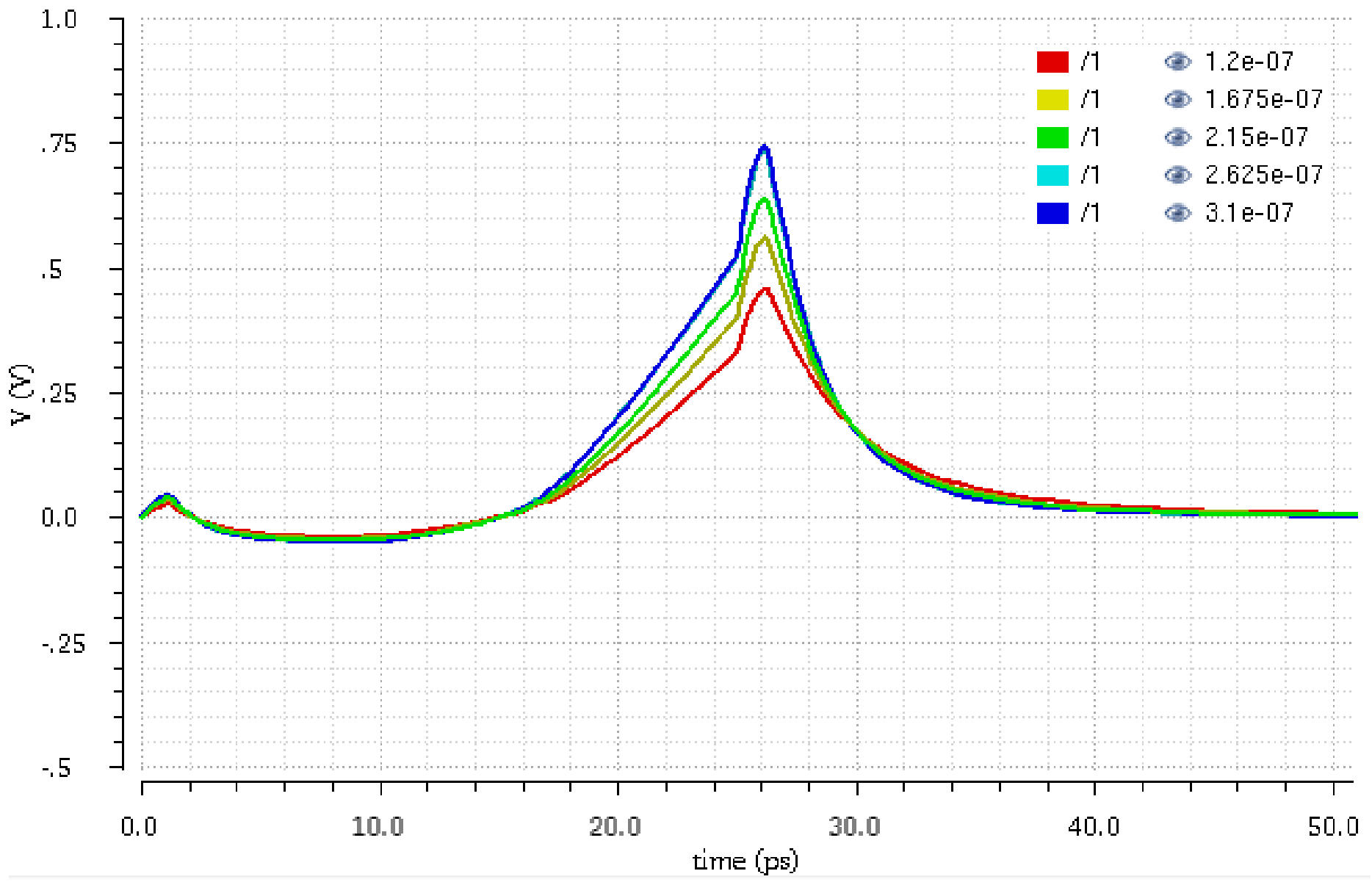

**Figure 4.** Parametric analysis of the MOSFET width

The proposed circuit is presented in Figure 3. It was implemented in the virtuoso analog design environment, utilising MOSFETs with a length of 100 nm. To observe the PFD's output response, a transient simulation setup with a 50 ps duration was created. Employing a parametric analysis, the MOSFET width was varied from 120 nm to 310 nm across 5 steps, as illustrated in Figure 4. The results demonstrate that the response curves for 310 nm and 262 nm overlap with each other, indicating their similar, swift responses. As the output curve 262 nm has the highest voltage, it indicates the fast-switching characteristics. Consequently, an operational frequency boost can be achieved with a 260 nm width, striking a balance between heightened performance and power consumption efficiency. After attaining the optimal length and width, the layout of the optimised PFD is crafted, which is presented in Figure 5.

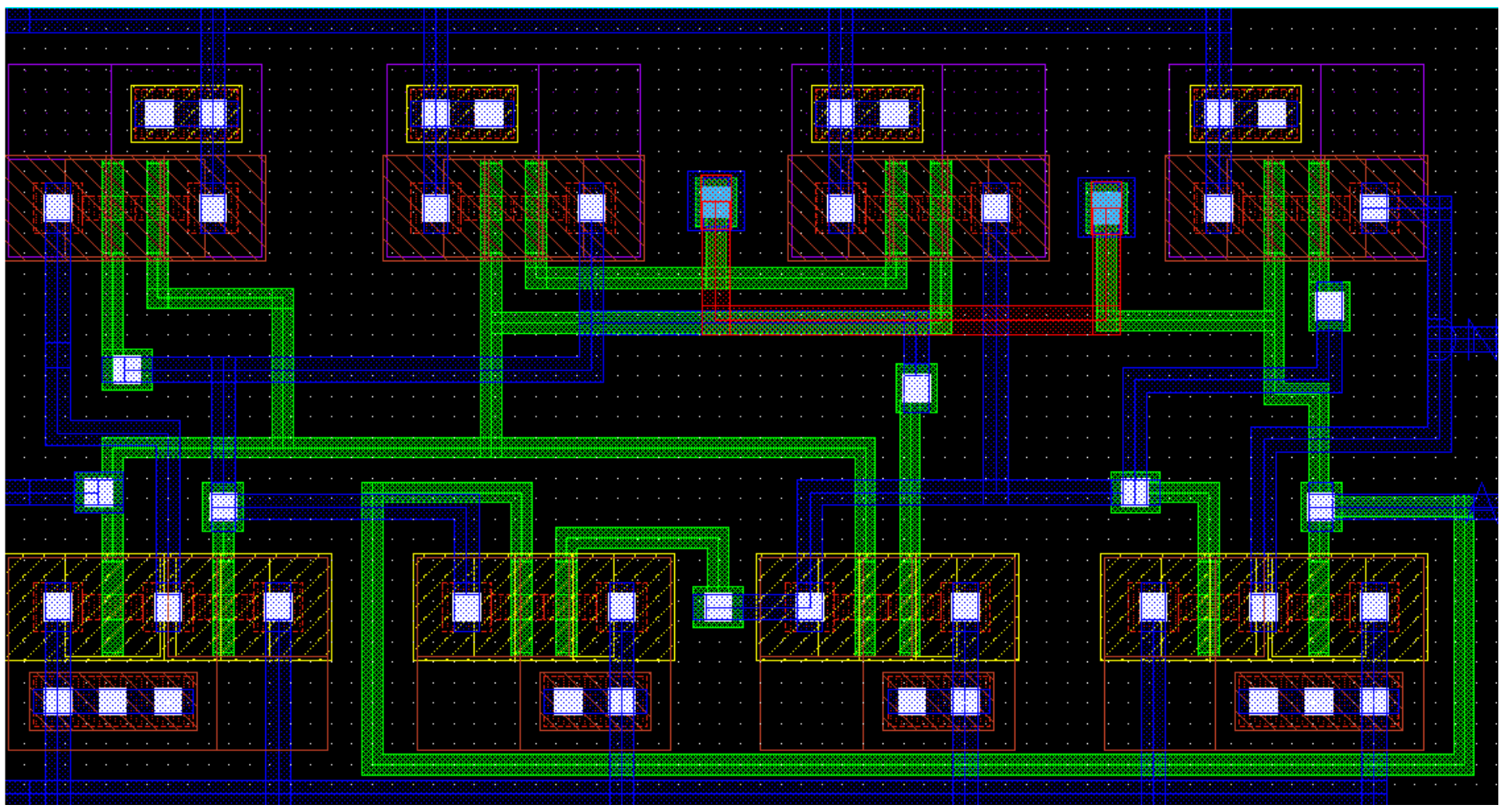

**Figure 5.** Layout design of the optimised PFD

## 3. Results and Discussions

The PFD is responsible for generating the UP and the Down signals according to the phase difference between the reference and the feedback signals. If the reference frequency leads the feedback signal, then UP signal becomes high. On the other hand, if the feedback frequency leads the reference frequency, the Down signal becomes high. In this section, the simulation results of the optimised symmetrical PFD are presented and analysed to evaluate its performance.

Figure 6 presents the transient analysis of the PFD with 1 GHz inputs with a 100 ps delay in feedback (presented as B) and reference (presented as A), respectively. From the figure, it can be observed that when the feedback frequency is delayed, the UP signal becomes high; whereas, if the reference frequency is delayed, the DOWN signal becomes high. These results match with the operational principles of the PFD. As a result, we can verify that, the proposed PFD is accurately functioning.

Within a PFD, the concept of dead zone holds critical significance. This term refers to a specific timeframe during when the PFD cannot differentiate between the feedback and reference frequencies. A larger dead zone leads to the introduction of unwanted jitter or noise in the device's operation. Figure 7 (a) illustrates the proper functioning of the PFD circuits, displaying an optimal delay of 25 ps in feedback. If the delay falls below this threshold, the PFD becomes incapable of distinguishing the disparity between the feedback and the reference frequencies, hampering its accuracy and reliability. From this simulation, we can verify the 25 ps dead zone of the proposed PFD.

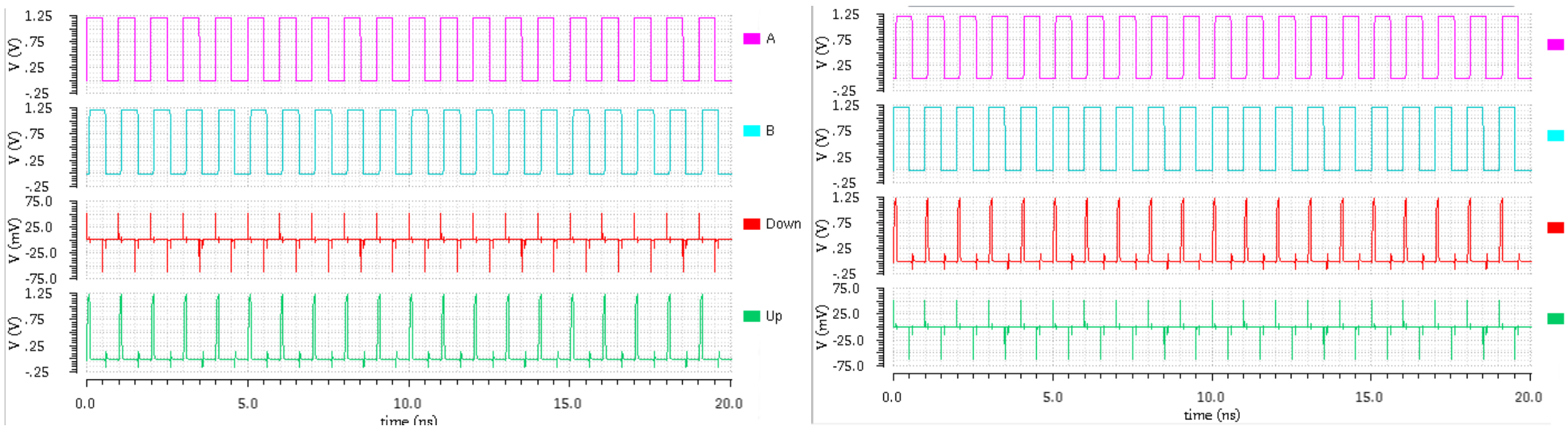

**Figure 6.** Transient analysis of PFD with a (a) feedback and a (b) reference delay of 100 ps

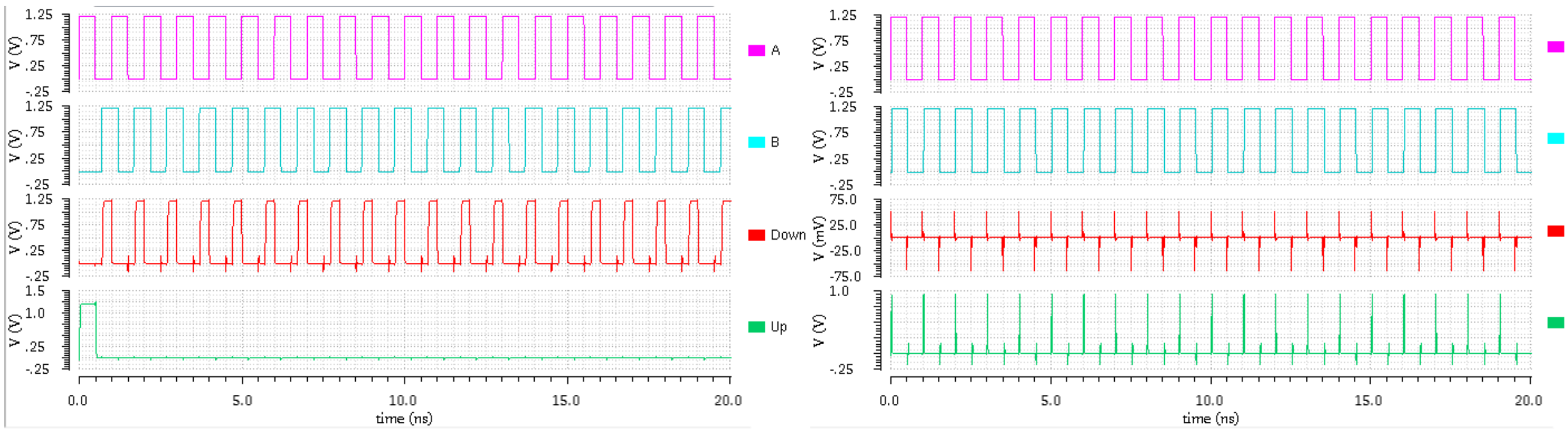

**Figure 7.** Transient analysis of PFD with feedback delay of (a) 25 ps and (b) greater than T/2

The PFD can behave differently in various conditions. In previous simulations, the phase difference between the input signals were small. Figure 7 (b) demonstrates the PFD's performance when there's a phase difference of T/2 between the reference and the feedback signals. Upon observation, it is evident that the PFD initially generates a UP signal and then smoothly stabilises, correctly operating thereafter.

It is essential to ensure that PFD operates correctly at higher frequencies., especially in the context of modern devices where swift data processing is paramount. Through the simulation presented in Figure 8 (a), it was established that the PFD effectively operates at a frequency of 5 GHz. Remarkably, the PFD demonstrated optimal performance within this high-frequency realm, affirming its suitability for contemporary, fast-paced data processing applications.

The proper functionality of the PFD across varying frequencies is compulsory. In Figure 8 (b), a transient analysis is presented, showcasing the scenarios where the reference and the feedback frequencies differ. Notably, the feedback frequency is observed to be lower than the reference frequency. Consequently, the PFD promptly generates a UP signal, indicating the need for an increase in the reference frequency for synchronisation.

Corner analysis in CMOS design refers to evaluating the circuit's performance under different process, voltage, and temperature (PVT) corners. These corners represent extreme conditions which the integrated circuits might encounter during their operation. By analysing how the CMOS circuit behaves under various corners, engineers can ensure its reliability and functionality across different scenarios. This analysis involves simulating the circuit's behaviour using worst-case conditions, such as minimum and maximum process variations, lowest and highest operating voltages and the broadest range of operating temperatures. Evaluating the circuit's performance under corner cases helps designers optimise the circuit for robustness and reliability, ensuring that the CMOS design can withstand real-world variations and consistently operate under varying environmental conditions. Figure 9 presents the corner analysis of the optimised PFD. The C1_0, C1_1, C1_2 and C1_3 in the graph in Figure 9 represents FF, FS, SF and SS corners, respectively. From the analysis we can verify that the PFD can accurately operate in various stress conditions.

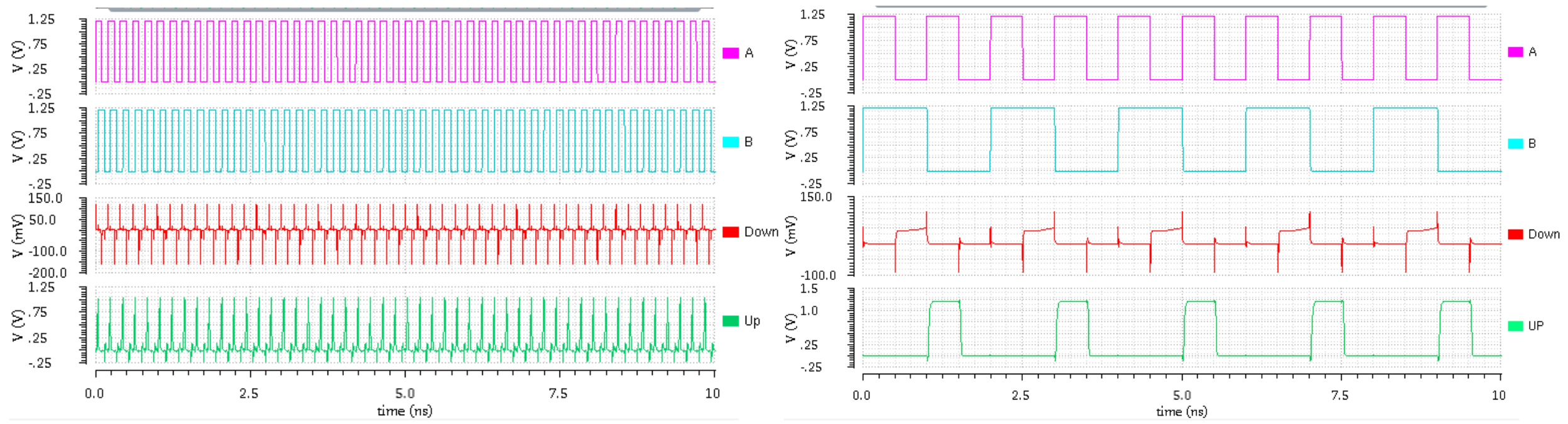

**Figure 8.** Transient analysis of the PFD with (a) 5GHz and (b) different frequencies.

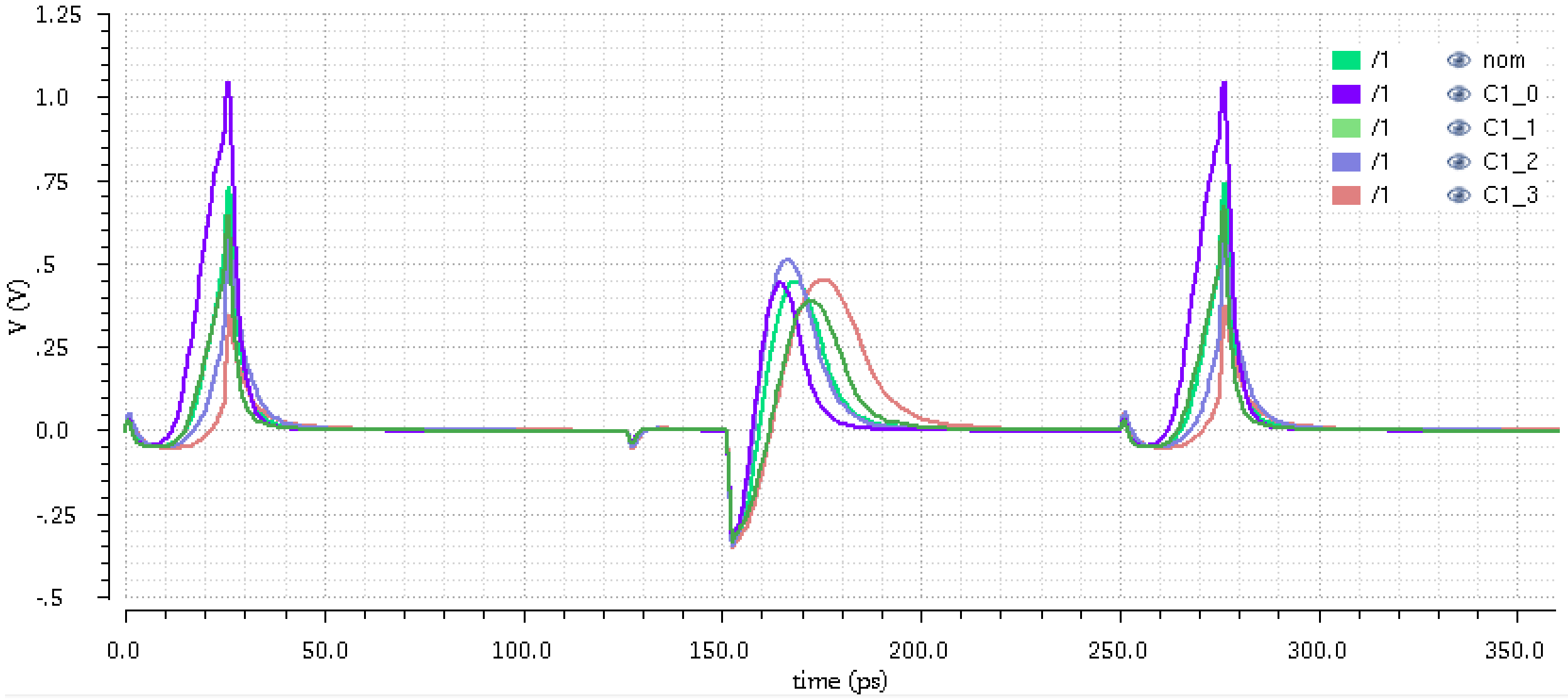

**Figure 9.** Corner analysis of the optimised PFD

**Table 1.** Performance parameters of the studied designs.

| Year | Topology | Process | Frequency | Dead zone | Power Consumption | Die Area |
|---|---|---|---|---|---|---|
| 2015 [14] | Pass Transistor And | IBM 130 nm | 0.75 GHz | 0.9 ns | 93.45 μW | |
| 2015 [14] | DIE Pass transistor PFD | IBM 130 nm | 0.9 GHz | 0.8 ns | 87.59 μW | |
| 2015 [14] | DIE Pass transistor NAND | IBM 130 nm | 1 GHz | 0.3 ns | 75.91 μW | |
| 2016 [15] | Transmission Gate based PFD | TSMC 130 nm | 1.7 GHz | 500 ps | 300 μW | |
| 2017 [16] | Modified conventional PFD | TSMC 130 nm | 3 GHz | 120 ps | 134 μW | |
| 2019 [17] | Symmetric PFD | TSMC 130 nm | 4.1 GHz | 25 ps | 90 μW | 250 μm$^2$ |
| 2019 [18] | Latch architecture-based PFD | 180 nm CMOS | 1 GHz | $\pi/11$ | 277 μW | |
| 2020 [19] | Modified true single-phase clock based PFD | 65 nm CMOS | 3.44 GHz | free | 324 μW | 322.612 μm$^2$ |
| 2020 [20] | Differential | TSMC 180 nm | 50 Hz -1GHz | $\pi/10$ | 129 μW | 225 μm$^2$ |
| 2020 [20] | Pseudo Differential | TSMC 180 nm | 50 Hz - 1GHz | $2\pi/25$ | 107 μW | 225 μm$^2$ |
| 2020 [21] | D flip-flop, TSPC logic, GDI logic-based PFD | TSMC 180 nm | 0.5 GHz - 3.33 GHz | free | 110 μW | |
| 2021 [22] | Precharged PFD | 180 nm CMOS | 1.25 MHz -3.8 GHz | free | 285.77 μW | |
| This Work | Optimised Symmetric PFD | 90 nm CMOS | 5 GHz | 25 ps | 29 μW | 28.431 μm$^2$ |

In Table 1, a detailed comparison of the performance of the recently optimised PFDs is presented. It has been established that the operational frequency of the PFD is intricately linked with the MOSFET's switching speed. By strategically increasing the MOSFET width, both the switching speed and the operational frequency were significantly enhanced. The other designs [17-22], covered in the above literature survey, are comparatively bulky, involving several stages of decision, which eventually decreased their operational frequencies. The results highlight a significant improvement in the proposed optimised design, showcasing a 5 GHz operational frequency, which is comparatively higher than those of the other designs. The optimised PFD also achieved reduced power consumption of only 29 μW, which is the lowest amongst the optimisation techniques, as shown in Table 1, due to the usage of only 16 MOSFETs. This indicates that the successful enhancement has been achieved through optimisation, underlining the efficiency and effectiveness of the optimised PFD configuration. The CMOS PFD design has a layout area of only 28.431 um2 which is also lowest among the studied designs. The compactness, power efficiency of the PFD gives the PFD and edge over other designs in terms of portability.

## 5. Conclusion

In summary, this study focused on optimising a PFD circuit by precisely optimising the PFD through advanced techniques involving the cadence virtuoso spectre parametric analysis and 90 nm CMOS technology. Through systematic parametric analysis, this research has optimised the PFD. Simulating the optimised symmetrical PFD under various circumstances, it has been observed that the optimised circuit exhibited an outstanding improvement, achieving a 0.9 GHz higher operational frequency, while consuming 61 μW less power compared to the other designs. The optimised PFD also achieved a reduced dead zone of only 25 ps. These results endorse the effectiveness of our approach in maximising operational efficiency, while minimising power consumption in PFD circuits. Even though the proposed design achieved higher frequency and low power consumption, the circuit had a dead zone of 25 ps. In the future, this can be further reduced to improve the PFD's performance drastically.

## Acknowledgement

This research is supported by the Ministry of Higher Education Malaysia [project code: FRGS/1/2020/TK0/XMU/02/5] and the Malaysia International Scholarship (KPT.B(S)700-4/2/2) scheme.